\documentclass[aps,prl,showpacs, twocolumn]{revtex4}

\usepackage{graphicx}
\usepackage{bm}
\usepackage{amsmath}

\providecommand{\bra}[1]{\langle #1 \rvert}
\providecommand{\ket}[1]{\lvert #1 \rangle}

\providecommand{\be}{\begin{equation}}
\providecommand{\ee}{\end{equation}}
\providecommand{\ba}{\begin{eqnarray}}
\providecommand{\ea}{\end{eqnarray}}

\usepackage{mathbbol}
\begin{document}

\title{Quantum search  with modular variables}
 
\author{A. Ketterer$^1$, T. Douce$^1$, A. Keller$^3$, T. Coudreau$^1$ and P. Milman$^1$}

\affiliation{$^{1}$Laboratoire Mat\'eriaux et Ph\'enom\`enes Quantiques, Universit\'e Paris Diderot, CNRS UMR 7162, 75013, Paris, France}
\affiliation{$^{3}$Univ. Paris-Sud 11, Institut de Sciences Mol\'eculaires d'Orsay (CNRS), B\^{a}timent 350--Campus d'Orsay, 91405 Orsay Cedex, France}

\begin{abstract}
We give a dimension independent formulation of the quantum search algorithm introduced in [L. K. Grover, Phys. Rev. Lett. {\bf 79}, 325 (1997)].  This algorithm provides a quadratic gain when compared to its classical counterpart by manipulating quantum two--level systems, qubits. We show that this gain, already known to be optimal, is preserved, irrespectively of the dimension of the system used to encode quantum information. This is shown by adapting the protocol to Hilbert spaces of any dimension using the same sequence of operations/logical gates as its original qubit formulation. Our results are detailed and illustrated for a system described by continuous variables, where qubits can be encoded in infinitely many distinct states using the modular variable formalism.
\end{abstract}
\vskip2pc 
 
\maketitle

\paragraph{Introduction} Using the mathematical structure behind quantum mechanics to design algorithms solving problems that can be stated classically can lead to striking advantages. The fascinating field of quantum algorithmic brought to light not only protocols over performing their classical analogs, but also provided to scientists in different fields a new way of looking at quantum mechanics. Along with the celebrated factorization quantum algorithm due to P. Shor \cite{Shor}, much attention has been paid to the optimal quantum search algorithm proposed by  L. K. Grover \cite{Grover1, Grover2}. Both algorithms have been experimentally realized in small scale in model systems \cite{ExpShor1, ExpShor2, ExpGrover1, ExpGrover2, ExpGrover3, ExpGrover4, ExpGrover5}.

Many other quantum algorithms have been proposed to solve different tasks \cite{site}, and the vast majority of them manipulate qubits \textit{i.e.}, quantum information encoded in the basis states of two dimensional Hilbert spaces. The question of whether higher dimensional quantum systems can be used to encode and process quantum information is an important one, both from the fundamental and the practical points of view: many experimental quantum systems are not ``naturally" two dimensional ones but are rather described by continuous variables (CV), as the electromagnetic field's quadrature or the position/momentum of a particle. Such systems can present advantages with respect to qubits from the point of view of decoherence (the main difficulty to the realization of a quantum computer) or detection efficiencies (that play a particular role in quantum communication protocols). Indeed, it is possible to define a set of universal quantum operations acting on quantum states described by CV \cite{Loyd, Loock}. Also, in the field of quantum communications, teleportation \cite{Kimble, Furosawa} and quantum key distribution \cite{Grosshans1, Grosshans2} protocols using CV were proposed and experimentally realized. In \cite{Gottesman} it was shown that qubits can be encoded in CV using some specific states consisting of infinite sums of infinitely squeezed states, and this idea is particularly suitable to the development of quantum error correcting codes. 

In the present manuscript, we use an original, more general, qubit encoding  \cite{Pierre, AndreasQC} to adapt to infinite dimensional Hilbert spaces described by CV the Grover search algorithm. In \cite{Pati}, a first adaptation of this algorithm was proposed for CV. The protocol developed here differs from this one in many aspects: first of all, the present protocol does not require experimentally challenging, infinitely squeezed states but rather uses arbitrary CV states to encode a qubit, as in \cite{AndreasQC}. As a consequence, the quantum search algorithm can be formulated using any CV product state as a starting point, as will be discussed below. Secondly, using the introduced formalism, the operations involved in the realization of the search protocol are the perfect CV analogs of the quantum gates realized in qubit systems. This is achieved by using a different way of looking at CV \cite{Pierre} using the manipulation of the modular variables formalism \cite{Aharanov}. Consequently, it is straightforward to determine the speed-up provided by the protocol and verify that it is the same one as in the qubit version of the algorithm. Finally, our protocol uses non gaussian, non unitary operations implemented by observables with a continuous spectrum but is, in spite of this fact, deterministic, as in \cite{Pati}. 

In order to demonstrate our results, we start by recalling the modular variables formalism \cite{Aharanov}. We consider the  dimensionless position and momentum like observables  $\hat{\theta}=2\pi \hat x/l$ and  $\hat k=l \hat p /\hbar$, where $l$ is an arbitrary length scale. Each observable can be split  into an integer and a modular part:
\begin{equation}
 \hat{\theta} =2\pi \hat N + \hat{\bar \theta} \qquad \hat k = \hat M + \hat{\bar k},
\label{eq:DefModVar}
\end{equation}
where $\hat N$ and $\hat M$ have integer eigenvalues and $\hat{\bar \theta}$ and $\hat{\bar k}$ are the dimensionless modular position and momentum operators with eigenvalues in the intervals $[0,2\pi[$ and $[0,1[$, respectively. 
Since $[\hat{\bar \theta}, \hat{\bar k}]=0$ \cite{Aharanov},   we can define a common  eigenbasis $\{\ket{\{\hat{\bar \theta},\hat{\bar k}\}}\}$, referred to as the \textit{modular basis}. Arbitrary quantum states in either position or momentum representation can be expressed in this basis, as $\ket{\Psi}=\int_0^{2\pi}\int_0^1 d\bar \theta d\bar k\ h({\bar \theta},{\bar k})\ \ket{\{\bar \theta,\bar k\}}$, with a complex normalized coefficient function $h:[0,2\pi[\times[0,1[\rightarrow \mathbb C$. Since the length scale $l$ is arbitrary, it can be chosen according to criteria as experimental convenience. 

As discussed in \cite{AndreasQC}, using the modular variables basis enables the definition of a continuum of two-level systems spanned by the states $\{\ket{\{\hat{\bar \theta},\hat{\bar k}\}},\ket{\{\hat{\bar \theta}+\pi,\hat{\bar k}\}}\}$ with which we can express the general state $\ket\Psi$ as:
\begin{align}
\ket{\Psi}=\int_0^{\pi}d\bar \theta \int_0^1 d\bar k f(\bar \theta,\bar k) \ket{\tilde{\Psi}(\bar \theta,\bar k)},
\label{eqn:GeneralState}
\end{align}
where now the $\bar{\theta}$ integral is over a $[0,\pi]$ interval and
\begin{align}
\ket{\tilde{\Psi}(\bar \theta,\bar k)}=&\cos{\alpha(\bar \theta,\bar k)} \ket{\{\bar \theta,\bar k\}} \nonumber \\
&+ \sin{\alpha(\bar \theta,\bar k)} e^{i\phi(\bar \theta,\bar k)} \ket{\{\bar \theta+\pi,\bar k\}}.
\label{eqn:continuousqubi}
\end{align} 
We also define a set of operators 
\begin{eqnarray}
\hat \Gamma_{\alpha} = \int_0^{\pi} d\bar \theta \int_0^1 d\bar k \  \zeta ({\bar \theta},{\bar k}) \hat \sigma_{\alpha}({\bar \theta},{\bar k}),
\label{eq:Gammas}
\end{eqnarray}
where $\alpha=x,y,z$, $\hat \sigma_{\alpha}({\bar \theta},{\bar k})$ are $SU(2)$ generators defined in each subspace $\{\ket{\{{\bar \theta},{\bar k}\}},\ket{\{{\bar \theta}+\pi,{\bar k}\}}\}$ and $\zeta({\bar \theta},{\bar k})$ is a real weight function \cite{comment2}, so that operators (\ref{eq:Gammas}) are hermitian. Operators (\ref{eq:Gammas}), that have a continuous spectrum, independently manipulate the $\{\bar \theta, \bar k\}$ dependent subspaces through the application of the $\{\bar \theta, \bar k\}$ dependent Pauli matrices \cite{AndreasQC, Pierre, comment2}. 

Of course, this {\it continuous discretization} can be applied also to multipartite CV systems. For a number $n$ of parties, we can define $n$ independent subspaces, denoted $\{{\bf   \boldsymbol{\bar \theta}}, {\bf \bar k}\}$ with $\boldsymbol{\bar \theta}=\bar \theta_1,...,\bar \theta_n $ and  ${\bf \bar k}=\bar k_1,...,\bar k_n $. Each of these subspaces can in turn be divided into continuums of two--level systems. One can also notice that the proposed decomposition of the CV space into a sum of two-dimensional subspaces can be equally well applied to finite discrete systems, showing the dimensional independency of the proposed approach. Nevertheless, in this manuscript, we will only discuss in detail the general case of CV, the case of finite discrete Hilbert spaces being a particular case of this one.

The general idea of the adaption of the Grover search algorithm to CV is to independently apply, to each one of the continuum of $\{{\bf   \boldsymbol{\bar \theta}}, {\bf \bar k}\}$ dependent  subspaces, the same sequence of operations as that used in the search algorithm defined for the usual qubits case. This can be done by using combinations of operators as (\ref{eq:Gammas}) such that, in each subspace, the usual qubit search algorithm is realized. We will see that this procedure ends up to be equivalent to performing a quantum search algorithm to qubits encoded in CV, and that the speed-up of the CV version of the algorithm is exactly the same as in the qubit case. 

\paragraph{The Grover search algorithm} We start by recalling the search  problem and the solution brought in \cite{Grover1,Grover2}. Assume one has a list of $N=2^n$ classical elements $s_i, i=1, \dots,N$, that can be encoded in a string of $n$ classical bits. We can  define a subset $\mathcal M$ of the list which contains $M<N$ elements which are searched. The search protocol can be formulated through the definition of a  function $f:\{i\}_{i=1,...,N} \rightarrow \{0,1\}$ for which $f(i)=1$ if $ s_i\in \mathcal M$ and $f(i)=0$ if $s_i \notin M$. Classically, applying this function to find the set  $\mathcal M$ takes ${\cal O}(N/M)$ calls to the function $f$. The quantum search algorithm introduced by L. Grover \cite{Grover1, Grover2} solves this classically stated problem  with a quadratic gain by using quantum mechanics to apply the function $f$ to a quantum list, that can be univocally defined from the classical one. As a result, the set  $\mathcal M$ is found in ${\cal O}(\sqrt{N/M})$ queries to the function $f$. 

An important point in Grover's solution is the definition of the encoding process, that defines  a quantum list from the classical one. In qubit terms, each element of the classical list can be encoded in different orthogonal states formed by the $N=2^n$ possible states of a  $n$-qubit system. The quantum list is thus defined as an equally weighted coherent superposition of all the orthogonal states (basis states) encoding the $s_i, i=1,...,N$ elements of the classical list. In practice, such a quantum list, described by the quantum state $\ket{\Psi_L}$, can be constructed using  the operator ${\bf H}=\text H^{\otimes N}$, the tensor product of a Hadamard gate applied to each qubit, such that $\ket{\Psi_L}={\bf H}\ket{0}^{\otimes N}$.

The quantum mechanical version of the search protocol thus consists of the repeated application of a so--called Grover operator,  $\hat G_\mathcal{M}$, which is defined univocally from the encoding process and the function $f$, determining $\mathcal{M}$. The complexity of the protocol is defined as the number of times the operator $\hat G_\mathcal{M}$ must be applied to the initial quantum list to find the searched elements with maximal probability. From the definition of the list, we can specify the operator $\hat G_\mathcal{M}=-{\bf H}{\cal I}_0{\bf H}{\cal I}_{\mathcal{M}}$, where ${\cal I}_{\mathcal{M}}=\mathbb 1-2\sum_{i\in \mathcal M}\ket{i}\bra{i}$ plays the role of the oracle (quantum operation determined by $f$) and  ${\cal I}_{0}=\mathbb 1-2(\ket{0}\bra{0})^{\otimes N}$ is a quantum phase gate. As an output of the quantum search algorithm, one obtains the searched state representing the set of searched classical strings, denoted as $\ket{s_f}=1/\sqrt{M}\sum_{i\in \mathcal M}\ket{i}$. Further on, we can infer the searched elements of $\mathcal M$ from $\ket{s_f}$. Also, using the state $\ket{\Psi_{\text L}}$ we can express the Grover operator as $G_{\mathcal{M}}={\cal I}_{\text L}{\cal I}_{\mathcal{M}}$, where ${\cal I}_{\text L}=(\mathbb 1-2 \ket{\Psi_{\text L}}\bra{\Psi_{\text L}})$.

\paragraph{From discrete to continuous variables} We now show how to adapt the described protocol to CV. For this, we consider the case where only one element is being searched, defining $\ket{s_f}=\ket{i}$. In this case,  the search operator can be denoted, in a simpler way, as $\hat G_\mathcal{M}=\hat G_{s_f}$. The starting point to convert the algorithm to CV is to  define operators $\hat G(s(\boldsymbol{\bar \theta},{\bf \bar k}))$,  that have exactly the same form as $\hat G_{s_f}$ but act independently on $\{{ \boldsymbol{\bar \theta}}, {\bf \bar k}\}$ dependent subspaces searching, in each one of them, an element denoted as $\ket{s(\boldsymbol{\bar \theta},{\bf \bar k})}$. The general idea  is as follows: multiple applications of the operators  $ \hat G(s(\boldsymbol{\bar \theta},{\bf \bar k}))$, corresponding to multiple oracle queries, realize independent quantum searches, as defined in the qubit realm, in each of the $\{{ \boldsymbol{\bar \theta}}, {\bf \bar k}\}$ subspaces. By doing so, it is possible to find one element in each one of the (equal dimension) $\{{ \boldsymbol{\bar \theta}}, {\bf \bar k}\}$ dependent subspace in a number of iterations that is exactly the same as the one in the usual qubit case. From this result, we will show that CV qubits can be defined as the integral of  all the ``found" elements in each one of the  $\{ \boldsymbol{\bar \theta}, {\bf \bar k}\}$ dependent subspaces. 

In order to better illustrate this general idea and the discrete to CV quantum search translation, we describe in detail the case of $N=2^2$ and $M=1$.  In this particular case, it is well known that the searched state $\ket{s_f}$, which is one among the $\ket{00}$, $\ket{01}$, $\ket{10}$ and $\ket{11}$ basis states is obtained  after {\it a single}  application only of the $\hat G_{s_f}$ operator. We will see how this same result can be obtained when qubits are encoded in CV. 

Using the continuous discretization introduced above, we start from an initial state encoded in CV  that is equivalent to $\ket{0}^{\otimes 2}$ in a two qubit system \cite{comment2}:
\begin{equation}\label{eq:list}
\ket{{\cal L}_0}=\int_0^{\pi}\int_0^1 {\bf d  \boldsymbol{\bar \theta}} {\bf d \bar k} g_1(\bar \theta_1, \bar k_1) g_2(\bar \theta_2, \bar k_2) \ket{\{\bar \theta_1, \bar k_1\}}\ket{\{\bar \theta_2,\bar k_2\}},
\end{equation} 
where $ {\bf d  \boldsymbol{\bar \theta}}= d\bar \theta_1 d\bar \theta_2$ , $  {\bf d \bar k}  =d \bar k_1 d \bar k_2$ and $g_i(\bar \theta_i, \bar k_i) $, $i=1,2$ are arbitrary functions defined in the domain $[0,\pi[ \times [0,1[$. As shown in \cite{AndreasQC}, state (\ref{eq:list}) can be seen as two qubit state encoded in CV: $\ket{{\cal L}_0}=\ket{\bar 0_1}\ket{\bar 0_2}$, with 
\begin{eqnarray}\label{CVqbit}
&&\ket{\bar 0_i}=\int_0^{\pi}\int_0^1 { d \bar \theta_i} { d \bar k_i} g_i(\bar \theta_i, \bar k_i)  \ket{\{\bar \theta_i, \bar k_i\}},
\end{eqnarray} 
$i=1,2$. A first remark is that we can compare Eqs. (\ref{eq:list}) and the definition (\ref{CVqbit}) to Eqs. (\ref{eqn:GeneralState}) and (\ref{eqn:continuousqubi}). By doing so, we can notice that, as in the case of usual qubits, the definition of the basis states in each $\{{ \boldsymbol{\bar \theta}}, {\bf \bar k}\}$ dependent subspace is arbitrary, and states (\ref{CVqbit}) were chosen for simplicity. Another point worth mentioning is that, given a  $\{{ \boldsymbol{\bar \theta}}, {\bf \bar k}\}$ dependent basis choice, the decomposition of the interval $[0,2\pi[$ in two domains appearing in (\ref{CVqbit}) is arbitrary as well. Thus, the definition of the $\ket{\bar 0}$ state is also arbitrary and can be encoded in any CV state. However, this choice will univocally determine the $\ket{\bar 1_i}$ states as the ones orthogonal to (\ref{CVqbit}):
\begin{eqnarray}\label{CVqbit1}
&&\ket{\bar 1_i}=\int_0^{\pi}\int_0^1 { d \bar \theta_i} { d \bar k_i} g_i(\bar \theta_i, \bar k_i)  \ket{\{\bar \theta_i+\pi, \bar k_i\}},
\end{eqnarray} 
In practice, the choice of decomposition and of the function $g_i(\bar \theta_i, \bar k_i)$, that can also be arbitrary, can be determined, for instance, by experimental possibilities and constraints. 

Further on, using (\ref{eq:Gammas}), we define a continuum of two qubit operators depending on $\{\bar \theta_1, \bar k_1\}$, $\{\bar \theta_2,\bar k_2\}$ that, combined, will lead to the definition of $\hat G(s(\boldsymbol{\bar \theta},{\bf \bar k}))$. We have, for instance, 
\begin{align}
{\boldsymbol {\cal H}}&( \boldsymbol{\bar \theta},{\bf \bar k}) ={ \cal H}(\bar \theta_1,\bar k_1) \otimes {\cal H}(\bar \theta_2,\bar k_2)  
\end{align}
where $\mathcal H(\bar \theta_{1(2)},\bar k_{1(2)})$ is the analog of the single qubit Hadamard operation, defined by:
\begin{align}
{\cal H}(\bar \theta_i,\bar k_i)) \ket{\{\bar \theta_i, \bar k_i\}}=\frac{1}{\sqrt 2} (\ket{\{\bar \theta_i, \bar k_i\}}+\ket{\{\bar \theta_i + \pi, \bar k_i\}}), \nonumber \\
{\cal H}(\bar \theta_i,\bar k_i)) \ket{\{\bar \theta_i+\pi, \bar k_i\}}=\frac{1}{\sqrt 2} (\ket{\{\bar \theta_i, \bar k_i\}}-\ket{\{\bar \theta_i + \pi, \bar k_i\}}),
\end{align}
$i=1,2$. We can also define
\begin{equation}\label{oracle}
{\cal I}_{s_f^i}( \boldsymbol{\bar \theta},{\bf \bar k})=\mathbb 1(\boldsymbol{\bar \theta},{\bf \bar k})-2\ket{s( \boldsymbol{\bar \theta},{\bf \bar k})}\bra{s( \boldsymbol{\bar \theta},{\bf \bar k})},
\end{equation}
where $\ket{s({\bf \boldsymbol{\bar \theta},\bar k})}=\ket{\{\bar \theta_1+\chi_{S_1}(\bar \theta_1)\pi, \bar k_1\}}\ket{\{\bar \theta_2+\chi_{S_2}( \bar \theta_2)\pi, \bar k_2\}}$ are the searched elements in each $\{\bar \theta_1, \bar k_1\}$, $\{\bar \theta_2,\bar k_2\}$ dependent subspace and {$\mathbb 1({\bf \boldsymbol{\bar \theta},\bar k})=\sum_{i,j=0,1}\ket{\{\bar \theta_1+i\pi}\ket{\{\bar \theta_2+j\pi} \bra{\{\bar \theta_1+i\pi}\bra{\{\bar \theta_2+j\pi}$}. The characteristic functions $\chi_{S_1}$ and $\chi_{S_2}$ defined by subsets $S_1$, $S_2 \subset[0,\pi]$ specify the searched item and are univocally determined by the function $f$.  From the chosen definition of states $\ket{\bar 0_{1(2)}}$ in (\ref{CVqbit}),  there are four possible choices of sets $S_{1,(2)}$:  $S_{1,(2)}= \emptyset$ or $S_{1,(2)}=[0,\pi]$   \cite{Tom}.

Finally, ${\cal I}_{0}( \boldsymbol{\bar \theta}, {\bf \bar k})=\mathbb 1(\boldsymbol{\bar \theta}, {\bf \bar k})-2\ket{\{\boldsymbol{\bar \theta}, {\bf \bar k}\}}\bra{\{\boldsymbol{\bar \theta}, {\bf \bar k})\}}$. 
Thus,  the $\{\boldsymbol{\bar \theta}, {\bf \bar k}\}$ dependent Grover operator can be defined as: 
\begin{eqnarray}
\hat G(s(\boldsymbol{\bar \theta},{\bf \bar k}))&=&-{\cal H}(\boldsymbol{\bar \theta},{\bf \bar k}) {\cal I}_{0}(\boldsymbol{\bar \theta},{\bf \bar k}) {\cal H}(\boldsymbol{\bar \theta},{\bf \bar k}) {\cal I}_{s}(\boldsymbol{\bar \theta},{\bf \bar k})\nonumber  \\
&=&{\cal I}_{\text L}(\boldsymbol{\bar \theta},{\bf \bar k})  {\cal I}_{s}(\boldsymbol{\bar \theta},{\bf \bar k}) 
\label{eqn:ContGroverOp}
\end{eqnarray}
with ${\cal I}_{\text L}(\boldsymbol{\bar \theta},{\bf \bar k})=\mathbb 1(\boldsymbol{\bar \theta},{\bf \bar k})-2\ket{\Psi_{\text L}(\boldsymbol{\bar \theta},{\bf \bar k})}\bra{\Psi_{\text L}(\boldsymbol{\bar \theta},{\bf \bar k})}$, $\ket{\Psi_{\text L}(\boldsymbol{\bar \theta},{\bf \bar k})}={\cal H}(\boldsymbol{\bar \theta},{\bf \bar k}) \ket{\{ \bar \theta_1,\bar k_1\}}\ket{\{ \bar \theta_2,\bar k_2\}}$. 

We can now use the $\{\bar \theta_1, \bar k_1\}$, $\{\bar \theta_2,\bar k_2\}$ dependent operators to define a bipartite continuous Grover operator by integrating $\hat G(s(\boldsymbol{\bar \theta},{\bf \bar k}))$ with real weight functions $\zeta_1(\bar \theta_1, \bar k_1)$ and $\zeta_2(\bar \theta_2, \bar k_2)$, as in (\ref{eq:Gammas}) \cite{comment2}:
\begin{equation}
\mathcal G_s=\int_0^{\pi}\int_0^{1} {\bf  d \boldsymbol{\bar \theta} d\bar k}  \zeta_1(\bar \theta_1, \bar k_1)\zeta_2(\bar \theta_2, \bar k_2)  \hat G(s(\boldsymbol{\bar \theta},{\bf \bar k})). 
\label{eq:ContGroverOp2}
\end{equation}
There are several points worth mentioning here. First of all, we can check that, in this case, operator (\ref{eq:ContGroverOp2}) has a continuous spectrum, and can be created by using operators as (\ref{eq:Gammas}) \cite{AndreasQC}. Secondly, it can be seen as the continuous limit of a multiple search protocol, as discussed above: by considering all the  $\{ \boldsymbol{\bar \theta},{\bf \bar k}\}$ independent subspaces as the whole list, and each one of the searched elements $\ket{s(\boldsymbol{\bar \theta},{\bf \bar k})}$ in each one of the subspaces as composing a set ${\cal M}$ of searched elements, we can easily check that this problem is equivalent to searching for one element in a list of four elements.  We can explore the consequences of these two points by applying (\ref{eq:ContGroverOp2}) to a continuous list, defined as 
\begin{equation}
\ket{\mathcal L}=\!\!\int_0^{\pi}\int_0^1 {\bf d\boldsymbol{\bar \theta}} {\bf d \bar k} g_1(\bar \theta_1, \bar k_1) g_2(\bar \theta_2, \bar k_2)  \ket{\Psi_{\text L} ({\bf \bar \theta,\bar k})}.
\label{eq:ContList}
\end{equation}
This list, as in the usual  formulation of the Grover algorithm,  can be obtained by the application of the operator $\boldsymbol {{\cal H}}=\int_0^1 \int_0^{\pi} d \boldsymbol{\bar \theta} {\bf d \bar k}  {\cal H}(\boldsymbol{\bar \theta},{\bf \bar k})$  to (\ref{eq:list}). 

Eq. (\ref{eq:ContList}) can be seen as the sum, with equal weights, of states:
\begin{subequations}
\begin{align}
&\ket{\bar 0_1}\ket{\bar 0_2}= \nonumber \\
&\int_0^{\pi}\int_0^1 {\bf d \boldsymbol{\bar \theta}} {\bf d \bar k}  g_1(\bar \theta_1, \bar k_1) g_2(\bar \theta_2, \bar k_2)   \ket{\{\bar \theta_1, \bar k_1\}}\ket{\{\bar \theta_2, \bar k_2\}},\label{eq:OrthState1} \\
&\ket{\bar 1_1}\ket{\bar 0_2}= \nonumber \\
&\int_0^{\pi}\int_0^{1}{\bf d \boldsymbol{\bar \theta}} {\bf d \bar k}  g_1(\bar \theta_1, \bar k_1) g_2(\bar \theta_2, \bar k_2)  \ket{\{\bar \theta_1+\pi, \bar k_1\}}\ket{\{\bar \theta_2, \bar k_2\}},\label{eq:OrthState2} \\
&\ket{\bar 0_1}\ket{\bar 1_2} =\nonumber \\
&\int_0^{\pi}\int_0^{1}{\bf d\boldsymbol{\bar \theta}} {\bf d \bar k}  g_1(\bar \theta_1, \bar k_1) g_2(\bar \theta_2, \bar k_2)  \ket{\{\bar \theta_1, \bar k_1\}}\ket{\{\bar \theta_2+\pi, \bar k_2}\}, \label{eq:OrthState3} \\
&\ket{\bar 1_1}\ket{\bar 1_2} =\nonumber \\
&\int_0^{\pi}\int_0^{1}{\bf d \boldsymbol{\bar \theta}} {\bf d \bar k}  g_1(\bar \theta_1, \bar k_1) g_2(\bar \theta_2, \bar k_2)  \ket{\{\bar \theta_1+\pi, \bar k_1\}}\ket{\{\bar \theta_2+\pi, \bar k_2\}}.\label{eq:OrthState4}
\end{align} 
\end{subequations}
Thus, equivalently, Eq. (\ref{eq:ContList}) is  the sum, with equal weights, of four possible CV encoded qubits, or four bipartite orthogonal states encoded in CV. 

An interesting point is that  Eq. (\ref{eq:ContList}) can be constructed from {\it any} CV state, as previously discussed, and that qubits in   (\ref{eq:OrthState1})-(\ref{eq:OrthState4}) can also be defined from arbitrary states. There is no need to rely on infinitely squeezed states: one can also use experimentally relevant states such as coherent or finite squeezed states.

In practice, operators (\ref{eq:ContGroverOp2}) and ${\cal H}$ can be realized using combinations of the $\hat \Gamma_{\alpha}$ operators, as shown in \cite{AndreasQC}.  After one run of the algorithm, consisting of one application of (\ref{eq:ContGroverOp2}) to (\ref{eq:ContList}), we obtain the final  state
\begin{align}\label{F}
\ket{{\cal F}_{s_i}}=\frac{1}{\sqrt{{\cal N}}}&\int_0^{\pi}\int_0^1 {\bf d \boldsymbol{\bar \theta}} {\bf d \bar k}  g_1(\bar \theta_1, \bar k_1) g_2(\bar \theta_2, \bar k_2)\nonumber \\
&\times  \zeta_1(\bar \theta_1, \bar k_1)\zeta_2(\bar \theta_2, \bar k_2) \ket{s ({\bf \bar \theta,\bar k})},
\end{align}
where $1/\sqrt{{\cal N}}$ is a normalization constant. At first sight, state (\ref{F}) (presumably the searched element) does not correspond to any of the four states  (\ref{eq:OrthState1})-(\ref{eq:OrthState4}) composing the initial quantum list $\ket{\mathcal L}$. Nevertheless, it can be  seen as one out of a set of four possible orthogonal states ($i=1,...,4$) that are characterized by the four  sets $S_{1,(2)}$, as is the case of states  (\ref{eq:OrthState1})-(\ref{eq:OrthState4}).  In other words, $\ket{{\cal F}_{s_i}}$  can be univocally associated to one of the four orthogonal states (\ref{eq:OrthState1})-(\ref{eq:OrthState4}) since it is non-orthogonal to only one of them.

We can also verify that
\begin{align}\label{eq:SumOrthStates}
\sum_{i=1}^4  \ket{{\cal F}_{s_i}}=\frac{1}{\sqrt{{\cal N}}}\int_0^{\pi}\int_0^{1} {\bf d \boldsymbol{\bar \theta}} {\bf d \bar k}  g_1(\bar \theta_1, \bar k_1) g_2(\bar \theta_2, \bar k_2)\nonumber \\
\times \zeta_1(\bar \theta_1, \bar k_1)\zeta_2(\bar \theta_2, \bar k_2) \ket{\Psi_L({\bf \bar \theta,\bar k})},
\end{align}
{\it i.e.}, apart from the presence of the multiplying functions $ \zeta_1(\bar \theta_1, \bar k_1)\zeta_2(\bar \theta_2, \bar k_2) $  the sum of possible search results  leads  to (\ref{eq:ContList}), analogously to the usual qubit version of the Grover algorithm.

The fact that, in general, the normalization constant ${\cal N} \neq 1$  (except in the case where $\zeta_1(\bar \theta_1, \bar k_1)=\zeta_2(\bar \theta_2, \bar k_2)=1$)   is a consequence of the non unitarity of the search operator (\ref{eq:ContGroverOp2}). However, this is by no means a limitation to the presented protocol for the following reasons: the ${\cal G}_s$ operators can be interpreted as part of a higher dimensional unitary operation, such that $\hat U_G={\cal G}_s\otimes {\cal R}+{\cal G'}_s\otimes {\cal R'}$ are operators acting on an auxiliary Hilbert space of dimension two or higher, and ${\cal G'}_s$ are operators as ${\cal G}_s$ with $\zeta_i( \bar \theta_i, \bar k_i) \rightarrow \zeta'_i(\bar \theta_i, \bar k_i) $ $i=1,2$ and ${\cal N}\rightarrow {\cal N}'$. Thus, operators ${\cal G'}_s$ also perform a quantum search where the obtained state is, as for ${\cal G}_s$,  univocally related to the searched state.

As in \cite{AndreasQC}, and in proposals for ancilla-based quantum computing \cite{AncillaDrivenQC}, the search algorithm can be interpreted as the application $\hat U_G$ to a composite quantum state consisting of  quantum states in the space where the quantum search is realized and an ancilla qubit initially prepared in state $\ket{0}$. Equivalently, the quantum search operators ${\cal G}_s$ and ${\cal G}'_s$ can be seen as  Kraus operators such that ${\cal G}_s^2+{{\cal G}'_s}^2=\mathbb 1$. 

We can easily find conditions such that the transformation  $\hat U_G$ is unitary.  A way of doing so, compatible with the definition of ${\cal G}_s$ and ${\cal G}'_s$, is to choose ${\cal R}=\hat \sigma_x$ and ${\cal R}'=\hat \sigma_z$. In this case, since the $\zeta_{1(2)}( \bar \theta_{1(2)}, \bar k_{1(2)})$ functions are real and the subspace dependent search operators $ \hat G(s(\boldsymbol{\bar \theta},{\bf \bar k}))$ satisfy $ \hat G^{\dagger}(s(\boldsymbol{\bar \theta},{\bf \bar k})) \hat G(s(\boldsymbol{\bar \theta'},{\bf \bar k'}))=\delta(\boldsymbol{\bar \theta'}-\boldsymbol{\bar \theta})\delta({\bf \bar k'}-{\bf \bar k} )\mathbb 1(\boldsymbol{\bar \theta},{\bf \bar k})$, we have that if  $\zeta_1^2( \bar \theta_1, \bar k_1)\zeta_2^2( \bar \theta_2, \bar k_2)+{\zeta'_1}^2( \bar \theta_1, \bar k_1){\zeta'_2}^2( \bar \theta_2, \bar k_2)=1$ then $U_G$ is unitary and:
\begin{equation}\label{U}
\hat U_G \ket{{\cal L}}\ket{0}= \ket{{\cal F}_{s_i}}\ket{0}+\ket{{\cal F'}_{s_i}}\ket{1}, 
\end{equation}
with $\bra{{\cal F}_{s_i}}{\cal G}_s^2\ket{{\cal F}_{s_i}}={\cal N}$ and $\bra{{\cal F'}_{s_i}}{\cal G'}_s^2\ket{{\cal F'}_{s_i}}={\cal N'}$. The important point in (\ref{U}) is that the two possible  search results $ \ket{{\cal F}_{s_i}}$ or $\ket{{\cal F'}_{s_i}}$ are, in fact, the same one after normalization, since both of them can be univocally associated to the same sets $S_{1(2)}$. 

The protocol and discussion above can be readily generalized to an arbitrary number of parties $n$, always following the same principles: an infinite sum of the  $\{\boldsymbol{\bar \theta},{\bf \bar k}\}$ operators with some  $\{\boldsymbol{\bar \theta},{\bf \bar k}\}$ dependent weight function or not (this fact doesn't change the protocol qualitatively) defined in each  $\{\boldsymbol{\bar \theta},{\bf \bar k}\}$ subspace is applied to an initial quantum list. This initial quantum list, as in the qubit case, is created by the application of the infinite sum of the Hadamard operator ${\cal H}=\int_0^{\pi}\int_0^{1} {\bf d \boldsymbol{\bar \theta}} {\bf d \bar k}{\cal H}(\boldsymbol{\bar \theta},{\bf \bar k})$ to an initial ``blank" state, that is the $n$ party analog of (\ref{eq:list}). Since our protocol can be seen, in each subspace, as the usual $n$ qubit Grover search algorithm, we obtain, after   ${\cal O}(\sqrt{n})$ applications of the search operator,  a state that can univocally associated to the searched element. 

\paragraph{Conclusion}

We have provided a systematic way to implement the Grover search algorithm to qubits encoded in CV. For this, we have shown that, in CV, any initial product state can be used as a ``quantum list" where well defined orthogonal quantum states can be identified. These discrete to continuous translation relies on the formalism of modular variables and deals with probabilistic operations that can be systematically interpreted so as to render the protocol deterministic. The observed overall gain obtained with the application of the protocol is exactly the same as the one obtained in the usual qubit formulation and the success of the proposed discrete to continuous variables quantum algorithm adaptation opens the perspective to building a common dimension independent formulation of quantum information. 

The authors thank S. P. Walborn for fruitful discussions.


\begin{thebibliography}{99}
\bibitem{comment2} Notice that we have the freedom of choice with functions $\zeta_i(\bar \theta_i, \bar k_i)$. It can be any function such that its absolute value is $\pi$ periodic and continuous at the edges of $\pi$ intervals.
\end{thebibliography}
\end{document}